\documentclass[aps,prd,nofootinbib,superscriptaddress,twocolumn]{revtex4}
\usepackage{graphicx}
\usepackage{amsmath}
\usepackage{amssymb}
\usepackage{amsfonts}
\usepackage{bbm}
\usepackage{amsbsy}
\usepackage{color}
\usepackage{cancel}
\usepackage{multirow}
\usepackage{soul}
\usepackage{marginnote}
\usepackage{tikz}
\usepackage{dsfont}
\usepackage{hyperref}
\newcommand{\ket}[1]{\ensuremath{\left|{#1}\right\rangle}}

\newcommand{\braket}[2]{\ensuremath{\langle{#1}|{#2}\rangle}}

\begin{document}

\title{Quantum-metrology estimation of spacetime parameters of the Earth outperforming classical precision}

\author{Jan Kohlrus}
\affiliation{School of Mathematical Sciences,
University of Nottingham,
University Park,
Nottingham NG7 2RD,
United Kingdom}

\author{David Edward Bruschi}
\affiliation{Faculty of Physics, University of Vienna, Boltzmanngasse 5, 1090 Wien, Austria}
\affiliation{Institute for Quantum Optics and Quantum Information - IQOQI Vienna, Boltzmanngasse 3, 1090 Vienna, Austria}

\author{Ivette Fuentes}
\affiliation{School of Mathematical Sciences,
University of Nottingham,
University Park,
Nottingham NG7 2RD,
United Kingdom}

\begin{abstract}
We consider quantum communication schemes where quantum optical signals are exchanged between a source on Earth and a satellite. The background curved spacetime affects the quantum state of the propagating photons. We employ quantum metrology techniques to obtain optimal bounds for the precision of quantum measurements of relevant physical parameters encoded in the final state. We focus on satellites in low Earth orbits and we find that our scheme improves the precision of the measurement of the Schwarzschild radius obtained within previous studies. Therefore, our techniques can provide the theoretical tools for novel developments that can potentially outperform the state-of-the-art obtained through classical means. We also review the impact of the relativistic effects on a simple quantum key distribution protocol within satellite schemes and find that such effects can be greatly damaging if they are not properly accounted for.
\end{abstract}
\maketitle

\section{Introduction}
Recent years have witnessed an increasing number of experimental proposals aimed at exploiting quantum systems to measure relativistic parameters, or aimed at operating in regimes where relativity plays a significant role. These proposals include experiments in the laboratory \cite{GIL} or schemes involving satellites \cite{satexpreview}. Quantum experiments with satellites have recently been incredibly successful: using the Micius satellite, quantum teleportations over 1400km have been demonstrated \cite{Pan_telep}, along with entanglement distributions \cite{Pan_entangl} and quantum key distribution (QKD) protocols \cite{Pan_QKD} established over 1200km. These breakthroughs have culminated with the deployment of an intercontinental quantum network between China and Austria \cite{Pan_network}. On the theoretical side, a body of work is now dedicated to the understanding of gravitational effects on space-based quantum science, with an eye on future tests within experiments and potential technological applications \cite{David1,David2,Kerr_sat}. This direction is still open, and many questions must be answered.

In the present work, we tackle the open question of the magnitude of relativistic effects on quantum states of photons that are exchanged between users at different heights in a gravitational field. We focus on the role of the kinematics of the satellite, and we show that the quantum state of photons that are reflected by a satellite back to an emitting ground station carries information of the reflection event. In particular, we show that the received photon states measured on Earth do not experience any gravitational or Doppler shift, but the change in the quantum state is mostly due to an energy kick delivered by the moving satellite's mirror. This energy shift is directly related the satellite's velocity, which itself depends on physical parameters such as the mass of the Earth and the radius of the satellite's orbit. We also consider previous link scenarios \cite{David1,David2,Kerr_sat}, but here without the constraint of using radial light beams. We employ quantum metrology techniques developed to study propagating photons in curved spacetime backgrounds \cite{David2,Kerr_sat} to compute the ultimate bound on precision measurements of relevant physical parameters of our basic setups. These parameters include the Schwarzschild radius of the Earth (or, equivalently, its mass) and the height of the satellite.

We find that the bounds on the relative error of measurements of the Schwarzschild radius can outperform the state-of-the-art. We note that our work is not optimized, therefore our result can be further improved if a thorough analysis of the optimal input states and final measurements is undertaken. 
We also look at the relativistic effects on QKD protocols set up using the same satellite schemes that we introduced for performing quantum metrology. We find that contrarily to the radial schemes used in \cite{David1,Kerr_sat}, QKD protocols using non radial beams of light are extremely compromised if relativistic effects are not properly corrected.
Our work shows the potential of aiding the proposal of novel tests of the overlap of relativity and quantum science, and the development of better performing relativistic quantum technologies.

This paper is organized as follows. In Sec. \ref{formalism}, we provide the mathematical tools for this work. In Sec. \ref{schemes}, we detail the satellite setups that we consider. In Sec. \ref{frequency_shifts} we give the energy shifts occurring for the photons in each scheme. In Sec. \ref{estimations_metrology} we employ quantum metrology techniques to estimate the bounds on precision measurements of physical parameters encoded in the states of propagating photons. An estimate of the magnitude of these effects for different orbits of the satellites can be found in Sec. \ref{values}. The effect on the quantum bit error rate (QBER) of a simple QKD protocol is given in Sec. \ref{QBER_section}. Discussions of our results can be found in Sec. \ref{conclusion}.

\section{Mathematical formalism}\label{formalism}
This section provides the mathematical machinery needed for our calculations. 
Throughout this work we use natural units $c=G=1$.

\subsection{Schwarzschild spacetime}\label{spacetime}
In this work we model the spacetime surrounding the Earth with Schwarzschild spacetime. This implies that we assume that the Earth is a static spherical mass distribution,  without any asperity or angular momentum. More refined spacetimes, such as the Hartle-Thorne \cite{Hartle-Thorne} or the Kerr \cite{Kerr} spacetime, can allow us to take into account such features. However, this comes at the price of much more involved analytical computations. In earlier work it was shown that the corrections due to the rotation of the Earth are at least two orders of magnitude lower than the Schwarzschild contributions \cite{Kerr_sat};
therefore, taking into account the rotation of the Earth will not affect the metrology results for static parameters such as the Schwarzschild radius in a way that can compete with the effects discussed in this work.
Given that the rotation of the Earth and its asphericity are negligible for the purposes of this work, they can be safely neglected.

The metric is a symmetric bilinear form which, in the usual Schwarzschild coordinates $(t,r,\theta,\phi)$ reads\footnote{Notice that this latter convention, which is widely used in physics, and where the polar angle $\theta=0$ denotes the North pole, is different from the geographic convention used for coordinates on our globe, where the latitude 0 is given by the equator.}
\begin{align}\label{metric}
\boldsymbol{g}=\text{diag}\left(-\left(1-\frac{2M}{r}\right),\frac{1}{1-\frac{2M}{r}},r^2,r^2 \sin^2{\theta}\right).
\end{align}
In natural units, the Earth's mass $M$ has the dimension of a length. The quantity $M/r$ is small, i.e., $M/r\ll1$, as can be seen by restoring units [we consider radii $r$ larger than the radius of the Earth $R_E$ and we have $GM/(R_E c^2)\ll1$]. 

\subsection{Observers in Schwarzschild spacetime}\label{observers}
In this work we consider different Earth-satellite schemes, which require us to introduce two types of observers: the observer on Earth, which is static, and (the observer comoving with) a satellite on a circular geodesic orbit orbiting around the Earth (i.e., contained in a plane intersecting the center of the sphere). These observers are defined by their motion, and their four-velocities are respectively
\begin{align}\label{static}
\boldsymbol{v_E} =&\, \frac{1}{\sqrt{1-\frac{2M}{R_E}}} \, \partial_t, \\ \label{sat}
\boldsymbol{v_s} =&\, \frac{\, \partial_t + \omega \, \partial_{\theta} + \zeta \, \partial_{\phi}}{\sqrt{1-\frac{2M}{R_s}-R_s^2(\omega^2+\zeta^2 \sin^2\theta_s)}},
\end{align}
where $R_s$ and $\theta_s$ are the radius of the orbit and the latitude coordinate of the satellite respectively. The angular coordinate velocities for the satellite are
\begin{align}\label{angularvel}
\omega =&\, \epsilon_{\omega} \sqrt{\frac{M}{R_s^3}} \sqrt{1-\frac{\sin^2{\alpha}}{\sin^2{\theta_s}}}, \quad \zeta = \epsilon_{\zeta} \sqrt{\frac{M}{R_s^3}} \frac{\sin{\alpha}}{\sin^2{\theta_s}}.
\end{align}
The angle $\alpha \in [0,\pi]$, with $\alpha\leq\theta_s$, measures the inclination of the satellite's plane of orbit with respect to the Earth's polar axis. The constants $\epsilon_{\omega}=\pm1$ and $\epsilon_{\zeta}=\pm1$ give the sign of the satellite's angular velocities (that is, the direction).

\subsection{Null vectors of light}\label{null_vector}
Here we characterize the null vectors that describe the paths of rays of light. This is important for our later computations.

\subsubsection{Characterization of the light ray}\label{null_vector_gen}
The trajectory followed by a light pulse that propagates in curved spacetime can be represented by a classical light ray. This light ray follows a null geodesic, which is defined by the field of null vectors tangent to it. In Schwarzschild spacetime, an arbitrary light ray has a null vector $\boldsymbol{k}$ of the form
\begin{align}\label{null}
\boldsymbol{k} =&\, E_p \Bigg( \frac{1}{1-\frac{2M}{r}} \, \partial_t + \epsilon_r \sqrt{1-\left(1-\frac{2M}{r}\right)\frac{l_{\phi}^2+\kappa}{r^2}} \, \partial_r \nonumber \\
&\, \quad \quad + \frac{\epsilon_{\theta}}{r} \sqrt{\frac{\kappa-l_{\phi}^2 \cot^2\theta}{r^2}} \, \partial_{\theta} + \frac{l_{\phi}}{r^2 \sin^2\theta} \, \partial_{\phi} \Bigg),
\end{align}
with $\epsilon_r=\pm1$ and $\epsilon_{\theta}=\pm1$ giving the sign of the radial and polar components of the null vector, respectively.
We have three constants of motion for each geodesic due to the spacetime being static and spherically symmetric. The first constant is $E_p$, which is the energy of the photon as it is measured by an inertial observer at space infinity. The constants $L_{\phi}$ and $K \geq 0$ are the azimuthal angular momentum and a quantity related to the square of the total angular momentum \cite{Carter} of the photon respectively, as seen by an inertial observer at spatial infinity. 
In \eqref{null}, we decided not to work with these two constants but we introduced the rescaled angular momentum and Carter constants $l_{\phi}=L_{\phi}/E_p$ and $\kappa=K/E_p^2$, in order to factor out the energy constant of the photon $E_p$. These two rescaled constants $l_{\phi}$ and $\kappa$ are independent on the energy of the photon. Notice that, from \eqref{null}, we can see that $\kappa \geq l_{\phi}^2 \cot^2\theta$.

\subsubsection{Nature of the conserved quantities}\label{conserved}
The energy constant of motion $E_p>0$ can be expressed in terms of the photon's frequency $\Omega$ measured by an observer with velocity $\boldsymbol{v}$ through the relation $\hbar \, \Omega = - \boldsymbol{k}\cdot\boldsymbol{v}$. For the static observer \eqref{static}, we thus have
\begin{align}\label{Ep_static}
E_p =&\, \hbar\,\Omega_E \sqrt{1-\frac{2M}{R_E}},
\end{align}
with the frequency of the photon $\Omega_E$ measured on Earth by the static observer \eqref{static}. 

Unfortunately, the two angular constants of motion $l_{\phi}$ and $\kappa$ cannot be obtained explicitly in a straightforward way. However, there are some special cases where one can obtain a simple relation between these two constants. For any radial light ray, namely with latitude $\theta=$ const and longitude $\phi=$ const, or for a trajectory constrained to the equatorial plane, we have $\kappa=0$. For a trajectory with a constant longitude $\phi$, we have $l_{\phi}=0$ and $\kappa \geq 0$. Radial photons are therefore obtained for $l_{\phi}=\kappa=0$.

The general procedure to obtain the constants $l_{\phi}$ and $\kappa$ is to integrate the relations $\frac{dr}{k^r}=\frac{d\theta}{k^{\theta}}$ and $\frac{d\theta}{k^{\theta}}=\frac{d\phi}{k^{\phi}}$, respectively. Given \eqref{null}, they read
{\small
\begin{align}\label{Carter_eq_gen_constraint}
\int_{r_1}^{r_2} \frac{\epsilon_r \, dr}{r^2 \sqrt{1-\left(1-\frac{2M}{r}\right) \frac{\kappa+l_{\phi}^2}{r^2}}} = \int_{\theta_1}^{\theta_2} \frac{\epsilon_{\theta} \, d \theta}{\sqrt{\kappa-l_{\phi}^2 \cot^2{\theta}}}, \\
l_{\phi} \int_{\theta_1}^{\theta_2} \frac{\epsilon_{\theta} \, d\theta}{\sin^2{\theta}\sqrt{\kappa-l_{\phi}^2 \cot^2{\theta}}}=\phi_2-\phi_1 \equiv \Delta \phi,\label{angular_cst_constraint}
\end{align}
}where the spatial coordinates $(r_1,\theta_1,\phi_1)$ and $(r_2,\theta_2,\phi_2)$ denote the starting and the ending points of the null geodesic, respectively. These coupled integral equations cannot be easily solved and we will make use of suitable approximations in order to obtain analytical insight.

\subsubsection{Light ray with small azimuthal angular momentum}\label{null_vector_per}
From now on we will consider light rays will a small azimuthal angular momentum $|l_{\phi}| \ll \sqrt{\kappa}$. In practice, this means that the longitude coordinate $\phi$ of the photon varies very slowly during the light's propagation. For such light rays, the null vector \eqref{null} simplifies to
\begin{align}\label{null_pert}
\boldsymbol{k} =&\, E_p \Bigg( \frac{1}{1-\frac{2M}{r}} \, \partial_t + \epsilon_r \sqrt{1-\left( 1-\frac{2M}{r} \right)\frac{\kappa}{r^2}} \, \partial_r \nonumber \\
&\, \quad \quad + \frac{\epsilon_{\theta}}{r} \sqrt{\frac{\kappa}{r^2}} \, \partial_{\theta} + \frac{l_{\phi}}{r^2 \sin^2\theta} \, \partial_{\phi} \Bigg) + O\left(\frac{l_{\phi}^2}{\kappa} \right).
\end{align}
The constraint \eqref{angular_cst_constraint} can now be easily integrated and gives
\begin{align}\label{angular_cst_constraint_pert}
\Delta \phi=&\, \epsilon_{\theta} \frac{l_{\phi}}{\sqrt{\kappa}} (\cot{\theta_1}-\cot{\theta_2}) + O\left(\frac{l_{\phi}^3}{\kappa^{3/2}} \right).
\end{align}
We therefore have
\begin{align}\label{azimuthal_cst_pert}
l_{\phi} \approx \sqrt{\kappa} \frac{\Delta \phi}{|\cot{\theta_2}-\cot{\theta_1}|}.
\end{align}
We emphasize that the condition holds when $|\Delta \phi| \ll |\cot{\theta_2}-\cot{\theta_1}|$.

Moving on to the constraint \eqref{Carter_eq_gen_constraint}, we see that it simplifies to
\begin{align}\label{Carter_eq_gen_constraint_pert}
\epsilon_r \int_{r_1}^{r_2} \frac{dr}{r^2 \sqrt{1-\left(1-\frac{2M}{r}\right) \frac{\kappa}{r^2}}} =&\, \frac{|\Delta \theta|}{\sqrt{\kappa}} + O\left(l_{\phi}^2 \right),
\end{align}
with $\Delta \theta \equiv \theta_2-\theta_1$. Note that the constraint doesn't depend on $l_{\phi}$ anymore. The integral in \eqref{Carter_eq_gen_constraint_pert} cannot be solved in a simple way, we will need to evaluate it numerically to obtain values of $\kappa$.

\subsection{Wave packets of photons}\label{quantum_formalism}
A photon can be modelled by a wavepacket, i.e., a superposition of states with different momentum and helicity \cite{helicity}. We assume that, for each momentum contribution to our initial state, there is a corresponding distribution of $\ket{+1}$ and $\ket{-1}$ helicity states with equal weights. Our full initial state then reads
\begin{align}\label{helicity superposition}
\ket{\boldsymbol{\gamma}}=\frac{1}{\sqrt{2}}\sum\limits_{s=\pm1} \int d\boldsymbol{p} \, F(\boldsymbol{p}) \ket{\boldsymbol{p},s},
\end{align}
with the helicity $s=\pm1$; the photon's four-momenta $\boldsymbol{p}=(k^{\hat{0}}, k^{\hat{1}}, k^{\hat{2}}, k^{\hat{3}})$ is measured in the observer's local Cartesian frame, and the distribution function $F(\boldsymbol{p})=F(\boldsymbol{\vec{p}}) \, \delta[|\boldsymbol{\vec{p}}|^2 -(k^{\hat{0}})^2] \, \bar{\theta}(k^{\hat{0}})$ of the four-momenta determines the weight of each state \ket{\boldsymbol{p}}. Furthermore, the function $\bar{\theta}$ is the step function and $F$ is the distribution of the three-momenta, which is normalized according to $\int d\boldsymbol{\vec{p}} \,\, |F(\boldsymbol{\vec{p}})|^2 = 1$. This also implies the normalization of the four-momenta distribution. As a consequence, the annihilation operators $\hat{a}_{\vec{p},s}$ for the photons have the usual bosonic commutation relations $[\hat{a}_{\vec{p}_1,s_1},\hat{a}_{\vec{p}_2,s_2}^{\dagger}]=\delta(\vec{p}_1-\vec{p}_2) \, \delta_{s_1s_2}$.

To further simplify the computations we assume that the observers pick a reference frame, implemented mathematically by a tetrad with components $e^{\, \, \, \mu}_{\hat{a}}$ (see, e.g., \cite{feliceclarke}), which is adapted to the prepared and received photons in the following way: the local momenta of the photons are $\boldsymbol{p}=(k^{\hat{0}}, 0, 0, -k^{\hat{0}})$. This choice can be easily achieved by picking $\boldsymbol{e}_{\hat{0}}=\boldsymbol{v}$ and $\boldsymbol{e}_{\hat{3}}=\frac{\boldsymbol{k}}{k^{\hat{0}}}-\boldsymbol{e}_{\hat{0}}$, while the first and second components of the triad $\boldsymbol{e}_{\hat{1}}$ and $\boldsymbol{e}_{\hat{2}}$ can be chosen arbitrarily (see \cite{Pa.Ta.We,helicity} for more details). The photon's energy $k^{\hat{0}}$, as seen in this local adapted frame at both emission and reception events is $k^{\hat{0}}=k^{\mu} e^{\hat{0}}_{\, \, \, \mu}=-k^{\mu} v^{\nu} g_{\mu \nu} = \hbar \, \Omega$, where $\Omega$ is the photon's frequency as seen by the observer with velocity $\boldsymbol{v}$. In the emitter's frame, the wavepackets have therefore the following expression:
\begin{align}\label{frequency_helicity_superposition}
\ket{\boldsymbol{\gamma}}=\frac{1}{\sqrt{2}}\sum\limits_{s=\pm1} \int d\Omega \, F(\Omega) \ket{\Omega, s}.
\end{align}

In the following, we consider schemes where the photons propagate between Earth and a satellite. We are thus interested in knowing what information about the propagation becomes encoded in the quantum state of the received photons. To this end, we note that momentum-helicity states of photons that propagate in a curved spacetime pick up a Wigner phase $\Psi$; see \cite{helicity}. If we denote by $\hat{\boldsymbol{U}}$ the operator that implements the desired propagation, we have
\begin{align}\label{phase factor}
\hat{\boldsymbol{U}} \ket{\Omega,s}=e^{i s \Psi(\boldsymbol{\vec{n}})} \ket{\Omega', s},
\end{align}
where the frequency $\Omega'$ of the state \textit{after} propagation in general in curved spacetime does not coincide with the emitted one, i.e., $\Omega' \neq \Omega$. We can relate these frequencies to each other by means of the frequency shift $f$ as $\Omega'=f \, \Omega$. Note that the Wigner phase $\Psi$ does not depend on the frequency of the photons however, it depends on their direction of propagation $\vec{n}=\vec{p}/|\vec{p}|$ as measured in the observer's local frame. After propagation the expression for the photon's state is
\begin{align}\label{helicity superposition2}
\ket{\boldsymbol{\gamma '}}=\int d\Omega \, F(\Omega) \ket{f \Omega} \otimes \sum\limits_{s=\pm1} \frac{e^{i s \Psi(\boldsymbol{\vec{n}})}}{\sqrt{2}} \ket{s}.
\end{align}
In expression \eqref{helicity superposition2} we made explicit the fact that the initial separable photon state \eqref{frequency_helicity_superposition} remains separable after propagation, as measured in the chosen adapted frames. Note that this is not true in general and this is solely due to our convenient choice of reference frame. Given a different choice of reference frame, wave packets of momenta would not depend on the frequency distribution only, and therefore the Wigner phase would depend on the variables of integration in \eqref{helicity superposition2}. This implies that the momentum and the helicity states would appear entangled due to the propagation in the curved spacetime. Although the measured state remains separable in our chosen frame, we emphasize that the Wigner phase is not a global phase. In particular, it has a different sign for each of the two helicity states of the superposition.

We can easily find a ``reference channel'', i.e., a channel where no Wigner phase is accrued, by employing wave packets with Bell-type helicity states of the form $(\ket{+-}+\ket{-+})/\sqrt{2}$, instead of the chosen qubit state $(\ket{+}+\ket{-})/\sqrt{2}$. Such rotationally invariant Bell states are superpositions of tensor products of opposite helicity eigenstates, which leads each vector in the state to acquire a total Wigner phase that vanishes. These states remain therefore invariant under relativistic effects \cite{invariantQI,Bradler}.

\section{Earth-satellite schemes}\label{schemes}
\begin{figure}[h!]
\includegraphics[width=\linewidth]{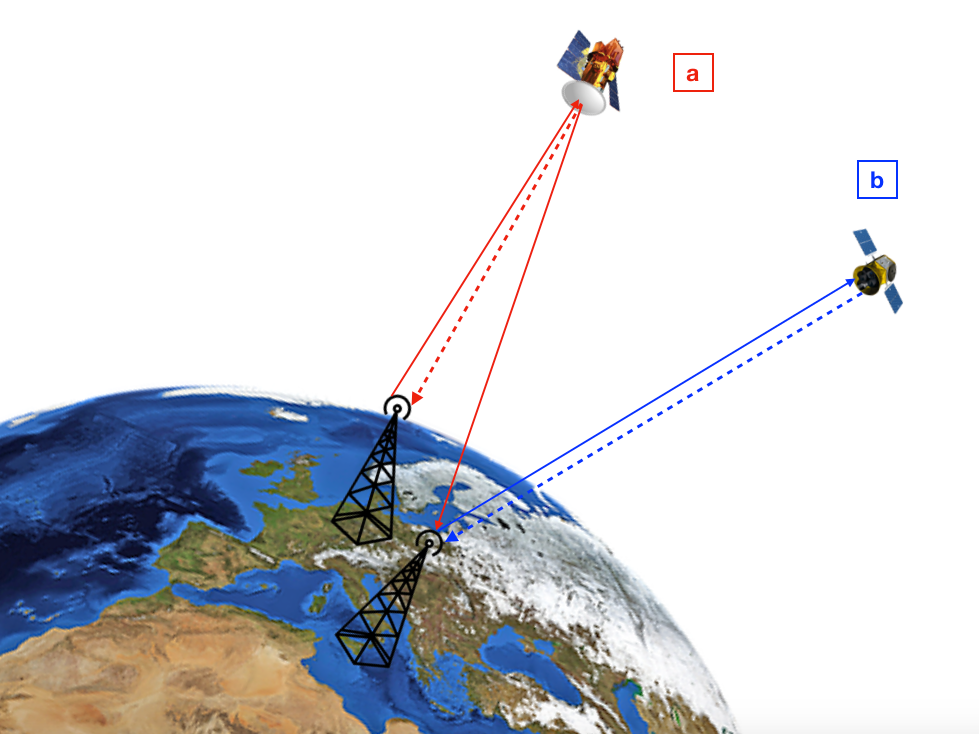}
\caption{Schematic representation of the proposed experiments. Scheme ``a'' (in red): photonic signals are sent from a station on Earth towards a satellite equipped with a mirror (in grey) and follow null geodesics (ascending thick line). They are reflected by the passing satellite (here placed at LEO) and they are then received at the same station (dotted line) or at another laboratory on Earth (descending thick line). Scheme ``b'' (in blue): quantum optical signals are sent from a station on Earth towards a satellite and follow null geodesics (thick line). Alternatively, a downlink can also be considered (dotted line).}
\label{figschemes}
\end{figure}

We now propose several schemes where photons are exchanged between a satellite and a station on Earth. Depending on the satellite considered, photons can be emitted from Earth to be reflected or detected by the satellite, or conversely the satellite can send photons towards Earth in a downlink. All these possible configurations are depicted in Fig. \ref{figschemes}.
The different schemes are described in more detail in the following subsections.

\subsection{Reflection scheme}\label{reflecting_scheme}

The setup of our first proposed scheme is the following and can be seen in Fig. \ref{figschemes} (scheme a, in red). A station on Earth emits photons towards a satellite equipped with a mirror, which is judiciously oriented in order to reflect the photons back to Earth to the same station or to another laboratory (see for example \cite{Villoresi1,Villoresi2,Villoresi3} for practical implementations). If the scheme involves two laboratories on Earth, for the sake of simplicity we assume that they are located almost at the same longitude $\phi$, although they can be arbitrarily far in terms of latitude $\theta$. 

Since the photons are prepared and measured by static observers at the same altitude, there will be no gravitational or Doppler shift. The frequency shift $f_r$ in this scheme is therefore only due to the change in energy that occurs at each reflection event on the satellites. 

This scheme is interesting in particular because the satellite is here only a passive object, which doesn't need to have any quantum source or measuring device on board.

\subsection{Link scheme}\label{link_scheme}

The second proposed scheme is similar to the schemes considered in the previous works \cite{David1,David2,Kerr}. It consists in a station on Earth exchanging photons with a satellite (see scheme b in Fig. \ref{figschemes}, in blue). However, here we do not restrain ourselves to radial light rays, and the photons can be sent from either the station or the satellite. This scheme can therefore be deployed in either uplink or downlink, and with angular light beams.

In this second scheme there is no reflection, therefore the frequency shift $f_l$ for this setup is made of the gravitational and Doppler shifts only. 

The main advantage of this scheme is that the photons need to cross the atmosphere only once. The atmosphere is a significant source of noise, the second scheme is therefore more resilient to such noise than the first one.

\subsection{Orbit configuration}\label{sat_orbit}

We now list different possible orbit configurations for the schemes described above.

	(i) $Orb1$. The satellite can be in geostationary orbit (GEO), which implies that its orbit lies in the Earth's equatorial plane, thus the inclination $\alpha=\theta_s=90^{\circ}$. 
	
	(ii) $Orb2$. We can also choose a satellite to be located in a low Earth orbit (LEO), which is an orbit at altitude $h\leq 2000$ km. 
	
	(iii) $Orb3$. The last family of orbits which we will consider are the very low Earth orbits (VLEO). They have an altitude $h\lesssim 300$ km, i.e., low in the thermosphere. Such satellites may experience significant atmospheric dragging due to the low altitude. However, the gravity field and steady-state ocean circulation explorer (GOCE) satellite orbited the Earth at $255$ km for more than four years due to its aerodynamic shape and ion propulsion \cite{GOCE}. This demonstrates that locating satellites at these altitudes is not impossible.

\section{Frequency shifts}\label{frequency_shifts}

In this section we describe how the frequencies of the photons are affected by the curved spacetime in each of the two schemes studied. In the reflection scheme, a change of energy is experienced by the photons at the reflection event. In the link scenario, gravitational redshift occurs gradually, all along the light's trajectory, until the detection where an extra Doppler shift contribution adds up due to the observer's motion.

\subsection{Reflection on satellite mirrors}\label{reflection}

Let us first describe the change induced at the reflection event occurring in the scheme described in Sec. \ref{reflecting_scheme}.

Both the null vector and the polarization vector of a light ray change when it is reflected by the satellite. The change in polarization may induce an extra Wigner phase contribution to the one acquired during propagation in \eqref{helicity superposition2}. Also, the momentum vector changes both in energy and direction: at the reflection event, some of the kinetic energy of the satellite is transferred to the photon and the light ray is deviated. We will see that the change in energy itself depends on how the light ray is deviated, since it depends on the satellite's motion and the ray's incident angle on the surface of the mirror.

The light ray with an arbitrary incident null vector $\boldsymbol{k}$ of the form \eqref{null} reflected by a moving satellite with velocity $\boldsymbol{v_s}$ given in \eqref{sat} will have a null vector $\boldsymbol{k'}$ of the form \eqref{null} as well, but with a different energy constant $E_p'$ and with new directional parameters $\epsilon_r'$, $\epsilon_{\theta}'$, $l_{\phi}'$, and $\kappa'$. The expression for the new energy constant $E_p'$ can be obtained by employing the constraint $\boldsymbol{k}\cdot\boldsymbol{v_s}=\boldsymbol{k'}\cdot\boldsymbol{v_s}$ at the reflection event. We find the reflection shift $f_r:=E_p'/E_p$ in energy to have the expression
\begin{align}\label{energy_reflect_exact}
f_r=&\, \frac{1-\sqrt{\frac{M}{R_s}} \delta_{\text{ang}}\left(\epsilon_{\theta},\kappa,l_{\phi}\right)
}{1-\sqrt{\frac{M}{R_s}}\delta_{\text{ang}}\left(\epsilon_{\theta}',\kappa ',l_{\phi}' \right)},
\end{align}
with the angular function {\small
\begin{align}\label{angular_function}
\delta_{\text{ang}}=&\, \epsilon_{\zeta} \, \frac{l_{\phi}}{R_s} \frac{\sin\alpha}{\sin^2\theta_s} + \epsilon_{\omega} \, \epsilon_{\theta} \sqrt{\frac{\kappa-l_{\phi}^2 \cot^2{\theta_s}}{R_s^2}} \sqrt{1-\frac{\sin^2\alpha}{\sin^2\theta_s}}.
\end{align}
}We introduced $\delta_{\text{ang}}$, a function encoding the angular parameters $\epsilon_{\theta}$, $\kappa$, and $l_{\phi}$ of the light ray. Numerical values for such changes in energy will be given later in Sec. \ref{param_values}. Notice that the change in energy \eqref{energy_reflect_exact} clearly encodes the change in the direction of the ray, apart from the sign of the radial direction $\epsilon_r$. If the photon's constants of propagation $\epsilon_{\theta}, \kappa$, and $l_{\phi}$ have the same value before and after the reflection, then $f_r=1$ and there is thus no effect. This occurs when the mirror's orientation is parallel to the satellite's three-velocity at the reflection event. This could be used to provide a reference channel defined as the channel with no effect, which could be used to account for any noise in the signal in scenarios where there is an effect. Also note that as long as the incident light ray is not radial the effect does not vanish if the light is reflected back to the initial laboratory. In this case, $\kappa'=\kappa$ but $\epsilon_{\theta}'=-\epsilon_{\theta}$ and $l_{\phi}'=-l_{\phi}$.

Let us now review how different satellite configurations affect the energy of the reflected photons.

$GEO$. For a reflection by a geostationary satellite, i.e., at latitude $\theta_s=\alpha=\pi/2$, the energy kick \eqref{energy_reflect_exact} reduces to:
\begin{align}\label{energy_reflect_GEO}
f_r=&\, \frac{1-\epsilon_{\zeta} \, \frac{l_{\phi}}{R_s} \sqrt{\frac{M}{R_s}}}{1-\epsilon_{\zeta} \, \frac{l_{\phi}'}{R_s} \sqrt{\frac{M}{R_s}}}.
\end{align}
This class of satellites only has an azimuthal velocity, which is the reason why the energy it can impart to the photons is only a function of the azimuthal part of the light's angular momentum. Since we will assume $l_{\phi}$ to be small, the change in energy due to the reflection by a GEO satellite can be expected to be small as well.

$LEO$. Another interesting class of satellites is those with polar orbits. These satellites pass above both poles during each revolution. Therefore, they are characterized by an inclination $\alpha=0$ and are typically found in LEO. The energy shift for this case is
\begin{align}\label{energy_reflect_LEO}
f_r=&\, \frac{1+\epsilon_{\omega} \, \epsilon_{\theta} \sqrt{\frac{\kappa \, M}{R_s^3}}}{1+\epsilon_{\omega} \, \epsilon_{\theta}' \sqrt{\frac{\kappa' M}{R_s^3}}} + O\left( \frac{l_{\phi}^2}{\kappa}\right).
\end{align}
Here we considered the approximation of small azimuthal angular momentum for the photons, that lead to expression \eqref{azimuthal_cst_pert}, namely $\kappa-l_{\phi}^2 \cot^2{\theta_s} \approx \kappa$, and therefore $\epsilon_{\theta} \sqrt{\kappa}$ represents the polar angular momentum constant of the photon. One can then notice the similarity of \eqref{energy_reflect_LEO} with \eqref{energy_reflect_GEO}. Yet, the effect here can be expected to be significantly higher since we do not make any assumption on the amplitude of the change in the latitude coordinate $\theta$ for the light ray.

\subsection{Gravitational and Doppler shifts}\label{grav_shift}

In the scenario described in Sec. \ref{link_scheme}, the photons with null vector $\boldsymbol{k}$ \eqref{null} emitted from Earth by the static observer with four-velocity $\boldsymbol{v_E}$ \eqref{static} to a satellite with velocity $\boldsymbol{v_s}$ \eqref{sat} experience a frequency shift $f_l$ of the form:
{\small
\begin{align}\label{freq_shift_exact}
f_l=&\, \left(1-\delta_{\text{ang}}\left(\epsilon_{\theta},\kappa,l_{\phi}\right) \sqrt{\frac{M}{R_s}}\right) \sqrt{\frac{1-\frac{2M}{R_E}}{1-\frac{3M}{R_s}}}.
\end{align}
}See Sec. \ref{param_values} for numerical values of such frequency shifts.
For a radial photon, namely when $\delta_{\text{ang}}$ vanishes, we recover the expression of the frequency shift from the previous works in Schwarzschild spacetime \cite{David1,David2}. Notice that the angular contribution to the shift comes in $\sqrt{M/R_s}\sim10^{-5}$, while the radial contribution comes in $M/R_s\sim10^{-10}$. That is, when the photons have sufficient angular momentum and thus non-negligible $\delta_{\text{ang}}$, the main contribution to the shift comes from this angular term. An interesting fact is that we can relate the frequency shift of the link scenario to the reflection shift of a light beam reflected radially:
\begin{align}\label{relation_shifts}
f_l=f_r\left(l_{\phi}'=\kappa '=0 \right) \, \sqrt{\frac{1-\frac{2M}{R_E}}{1-\frac{3M}{R_s}}}.
\end{align}

Let us now study the frequency shifts experienced by photons for specific classes of satellites orbits.

$GEO$. For satellites orbiting in the Earth's equatorial plane, the frequency shift reduces to
\begin{align}\label{freq_shift_GEO}
f_l=&\, \left(1-\epsilon_{\zeta} \frac{l_{\phi}}{R_s} \sqrt{\frac{M}{R_s}}\right) \sqrt{\frac{1-\frac{2M}{R_E}}{1-\frac{3M}{R_s}}}.
\end{align}
Here again, as we consider photons with small azimuthal angular momenta, this shift is likely to be small.

$LEO$. For low altitude satellites that we consider to follow polar orbits, we have
\begin{align}\label{freq_shift_LEO}
f_l=&\, \left(1-\epsilon_{\theta} \, \epsilon_{\omega} \sqrt{\frac{ \kappa \, M}{R_s^3}}\right) \sqrt{\frac{1-\frac{2M}{R_E}}{1-\frac{3M}{R_s}}} + O\left( \frac{l_{\phi}^2}{\kappa}\right).
\end{align}

\section{Quantum metrology of physical parameters of the Earth}\label{values_metrology}
In this section, we use quantum-metrology techniques to derive the optimal bounds on precision measurements of spacetime parameters of the Earth. These parameters are encoded in the states \eqref{helicity superposition2} that are received after following one of the paths described above.
We first compute a theoretical bound and then provide numerical values in order to compare the magnitude with the state-of-the-art.

\subsection{Quantum metrology of spacetime parameters}\label{estimations_metrology}

Quantum metrology is a theory that provides tools to estimate parameters encoded in quantum states that evolve through a unitary channel \cite{QFI}. The channel can model a variety of different physical scenarios which are described by a parameter such as time, temperature, or a small perturbation induced by changes in the spacetime curvature \cite{rqm1,rqm2}. The final state therefore encodes some information about the parameter, and it can be measured in order to extract such information. In general, the precision that can be achieved is bounded by the laws of quantum mechanics, and it depends on the input state and the final measurement \cite{QFI}.

Given a unitary channel $\hat{U}_\epsilon$ parametrized by the parameter $\epsilon$ to be measured, it is possible to show that the quantity $H(\epsilon)$ known as quantum Fisher information bounds the precision $\Delta\epsilon$ on the measurement through the Cramer-R\`ao bound $(\Delta\epsilon)^2\geq1/[N\,H(\epsilon)]$, where $N$ is the number of input probes \cite{Cramer}.

We will use this bound in our estimation of the precision of measurement schemes using the setups proposed in this work.

\subsubsection{Overlap between emitted and measured states}\label{overlap_subsect}
In our present study, the initial state of the photons is \eqref{frequency_helicity_superposition}, while the received state \eqref{helicity superposition2} picks up a Wigner phase $\Psi$ and an energy shift after the pulse of light has traveled through its path in space. The overlap $\Theta=\braket{\boldsymbol{\gamma}}{\boldsymbol{\gamma'}}$ between the emitted and received states can be easily computed and reads
\begin{align}
\Theta=&\, \cos{\left(\Psi\right)} \int d\Omega \, F^*(f \Omega) F(\Omega), \label{overlap_result}
\end{align}
which has required us to use the relations $\braket{\tilde{s}}{s}=\delta_{\tilde{s}s}$ and $\braket{\tilde{\Omega}}{f \Omega}=\delta(\tilde{\Omega}-f \Omega)$.
Let us consider, as in previous studies \cite{David1,David2,Kerr_sat}, a real Gaussian distribution $F$ of the form
\begin{align}\label{gaussian}
F(\Omega)=\frac{1}{(2\pi \sigma^2)^{\frac{1}{4}}} \exp{\left(-\frac{(\Omega-\Omega_0)^2}{4\sigma^2}\right)}.
\end{align}
We are interested in obtaining an explicit expression for the momentum contribution $I:=\, \int d\Omega \, F^*(f \Omega) F(\Omega)$ of the overlap \eqref{overlap_result}.
Using \eqref{gaussian} together with \eqref{energy_reflect_exact}, we obtain
{\small
\begin{align}\label{momentum_overlap_reflect}
I \approx&\, \exp\left[-\left(\delta_{\text{ang}}\left(\epsilon_{\theta},\kappa,l_{\phi}\right)-\delta_{\text{ang}}\left(\epsilon_{\theta}',\kappa ',l_{\phi}' \right) \right)^2 \frac{\Omega_0^2}{8 \sigma^2}\epsilon^2\right].
\end{align}
}In the above expression we have introduced the dimensionless perturbative parameter $\epsilon=\sqrt{M/R_s} \sim 10^{-5}$, and we have assumed that we work in a regime where $\epsilon \ll \epsilon^2 (\Omega_0^2/\sigma^2)$. From now on we will only give expressions in this limit.

In this regime the Wigner phase is small \cite{helicity} and the effects computed here are significantly larger. Therefore, from now on we will ignore the contribution from the effects of gravity on the helicity of the photons. Note that \eqref{momentum_overlap_reflect} was derived for the reflecting scheme in Sec. \ref{reflecting_scheme}, using \eqref{energy_reflect_exact}. However, the expression of the overlap for the link scheme described in Sec. \ref{link_scheme} is
\begin{align}\label{momentum_overlap_link}
I \approx&\, \exp\left(-\delta_{\text{ang}}^2\left(\epsilon_{\theta},\kappa,l_{\phi}\right) \frac{\Omega_0^2}{8 \sigma^2}\epsilon^2\right).
\end{align}
Namely, we can use \eqref{momentum_overlap_reflect} for both schemes provided that we set $\delta_{\text{ang}}\left(\epsilon_{\theta}',\kappa ',l_{\phi}' \right)=0$ for the link scheme.

\subsubsection{Fidelity of quantum states and quantum Fisher information}\label{metrology}
The fundamental quantity needed to estimate the precision of quantum measurements of spacetime parameters is the quantum Fisher information (QFI) \cite{QFI}. The QFI is expressed in terms of the fidelity $\mathcal{F}(\ket{\boldsymbol{\gamma_{\epsilon}}},\ket{\boldsymbol{\gamma_{\epsilon+d\epsilon}}})$ between two neighboring states in the Hilbert space, $\ket{\boldsymbol{\gamma_{\epsilon}}}$ and $\ket{\boldsymbol{\gamma_{\epsilon+d\epsilon}}}$, that encode the same parameter $\epsilon$ and are infinitesimally apart (i.e., $d\epsilon\ll1$). We are not interested here in the full exposition of the  theory but will provide only the main steps, leaving the details of the theory of relativistic quantum metrology to more in depth literature \cite{rqm1,rqm2}.

We start by computing the fidelity between the received state that encodes the spacetime's parameter $\epsilon$ and the fiducial emitted one. The latter has a similar expression than the former but with $\epsilon=0$. The fidelity between these two quantum states is simply $\mathcal{F}(\ket{\boldsymbol{\gamma_{0}}},\ket{\boldsymbol{\gamma_{\epsilon}}})=|\Theta|^2\approx I^2$. Let's now consider two states that have both gone through the path of one of the schemes described in Sec. \ref{schemes}, but that have experienced an energy shift with an infinitesimal difference $d\epsilon$ in the parameter $\epsilon=\sqrt{M/R_s}$.
We use the fidelity between these two states to define the QFI $H(\epsilon)$ as
\begin{align}\label{QFI_def}
H(\epsilon) = \lim_{d\epsilon \to 0} 8 \, \frac{1-\sqrt{\mathcal{F}(\ket{\boldsymbol{\gamma_{\epsilon}}},\ket{\boldsymbol{\gamma_{\epsilon+d\epsilon}}})}}{d \epsilon^2}.
\end{align}
We can then obtain the expression for the $\epsilon=0$ contribution of the QFI, $H^{(0)} =\, \lim_{d\epsilon \to 0} 8 \, \frac{1-\sqrt{\mathcal{F}(\ket{\boldsymbol{\gamma_{0}}},\ket{\boldsymbol{\gamma_{d\epsilon}}})}}{d \epsilon^2} $, which reads, in the regime we consider,
\begin{align}\label{QFI0_expl}
H^{(0)} \approx&\, \left[\delta_{\text{ang}}\left(\epsilon_{\theta},\kappa,l_{\phi}\right)-\delta_{\text{ang}}\left(\epsilon_{\theta}',\kappa ',l_{\phi}' \right) \right]^2 \frac{\Omega_0^2}{\sigma^2}.
\end{align}
To obtain \eqref{QFI0_expl}, we used the expression of \eqref{momentum_overlap_reflect} for an infinitesimal value $d\epsilon$ of our parameter $\epsilon$. We could then expand the expression up to lowest order in $d\epsilon \frac{\Omega_0}{\sigma} \ll 1$. Inserting this expression into \eqref{QFI_def} together with $\epsilon=0$, and finally taking the limit $d\epsilon\to0$, we obtain the result above.

The measurement precision of a parameter $\epsilon$ encoded in the received quantum states is given to lowest order by the Cram\'er-Rao inequality \cite{Cramer} as
\begin{align}\label{Cramer-Rao}
|\Delta \epsilon| \geq \frac{1}{\sqrt{N H(\epsilon)}}  \approx \frac{1}{\sqrt{N H^{(0)}}},
\end{align}
where $N$ is the number of probes used in the experiment. We are here able to work with the lowest-order term of the QFI since it is nonzero. Higher-order terms from the QFI \eqref{QFI_def} would be negligible in the regime $\epsilon \ll \epsilon^2 (\Omega_0^2/\sigma^2)$ we are working in. We now have all the tools required to give the bound on precision of physical measurement of spacetime parameters of the Earth.

\subsubsection{Estimation of physical parameters of the Earth}\label{estimations}
Our main control parameter $\epsilon=\sqrt{M/R_s}$ is a function of physically relevant parameters. For example, we can use it to estimate the bound for the precision on the Schwarzschild radius of the Earth $r_S=2M$. First we note that $|\Delta \epsilon| = \frac{|\Delta M|}{2 \sqrt{M R_s}}=\frac{|\Delta r_S|}{2 \sqrt{2 r_S R_s}}$, which then allows us to obtain
\begin{align}\label{rs_bound}
\frac{|\Delta r_S|}{r_S} \geq&\, \frac{2 \sqrt{2}}{\sqrt{N H^{(0)}}} \sqrt{\frac{R_s}{r_S}}.
\end{align}
The other parameter encoded in $\epsilon$ is the satellite's orbital radius $R_s=R_E+h$. We find $|\Delta \epsilon| =\, \frac{|\Delta R_E|}{2(R_E+h)} \sqrt{\frac{M}{R_E+h}}$ and $|\Delta \epsilon| =\, \frac{|\Delta h|}{2(R_E+h)} \sqrt{\frac{M}{R_E+h}}$, which give us the following bounds for the precision on the Earth's local radius and the satellite's altitude
\begin{align}\label{RE_bound}
\frac{|\Delta R_E|}{R_E} \geq&\, \frac{2}{\sqrt{N H^{(0)}}} \frac{R_s}{R_E} \sqrt{\frac{R_s}{M}}, \\ \label{h_bound}
\frac{|\Delta h|}{h} \geq&\, \frac{2}{\sqrt{N H^{(0)}}} \frac{R_s}{h} \sqrt{\frac{R_s}{M}}.
\end{align}

\subsection{Numerical values}\label{values}

\subsubsection{Numerical values of the parameters}\label{param_values}
In the expressions of the energy shifts \eqref{energy_reflect_exact} and \eqref{freq_shift_exact}, there are physical parameters of the Earth, of the light's trajectory and of the satellite's orbit. The parameters of the Earth are: its (mean) radius $R_E=6371$ km, and its mass which takes the value of $M=4.43$ mm in natural units. In addition, the satellite orbits are located at altitude $h=R_s-R_E$. For LEO orbits, we choose a polar orbit with inclination parameter $\alpha=0^{\circ}$, and with $h = 2000$ km, namely the highest  altitude to be considered as LEO. For the VLEO ones, we take the parameters of the GOCE satellite, i.e., altitude of $h =255$ km and inclination of $\alpha=6.7^{\circ}$. In Table \ref{paramtable}, we give the values for the parameters appearing in the expressions of the energy shifts, together with the corresponding shifts.

\begin{table}[!htbp]
\begin{tabular}{| c | c | c | c |}
\hline
\multirow{2}{*}{Quantity} & \multirow{2}{*}{Orbit} & \multirow{2}{*}{Light's trajectory} & \multirow{2}{*}{Value} \\ 
 & & & \\ \hline \hline
\multirow{2}{*}{$M/R_s$} & LEO & / & $5.29 \times 10^{-10}$ \\  
\cline{2-4}
 & VLEO & / & $ 6.69 \times 10^{-10}$ \\ \hline
 \multirow{4}{*}{$l_{\phi}/R_s$} & \multirow{2}{*}{LEO} & Sat $\rightleftarrows$ Lab1 & $\pm 1.01\times 10^{-4}$ \\ \cline{3-4}
 & & Sat $\rightleftarrows$ Lab2 & $\mp 8.99\times 10^{-5}$ \\ \cline{2-4}
 & VLEO & Sat $\rightleftarrows$ Lab1 & $\pm 7.64 \times 10^{-4}$ \\ \hline
  \multirow{4}{*}{$\kappa/R_s^2$} & \multirow{2}{*}{LEO} & Sat $\rightleftarrows$ Lab1 &  0.49 \\ \cline{3-4}
 & & Sat $\rightleftarrows$ Lab2 & 0.58 \\ \cline{2-4}
 & VLEO & Sat $\rightleftarrows$ Lab1 & 0.88 \\ \hline
   \multirow{4}{*}{$\delta_{\text{ang}}$} & \multirow{2}{*}{LEO} & Sat $\rightleftarrows$ Lab1 & $\mp 0.70$ \\ \cline{3-4}
 & & Sat $\rightleftarrows$ Lab2 & $\mp 0.76$ \\ \cline{2-4}
 & VLEO & Sat $\rightleftarrows$ Lab1 & $\mp 0.91$ \\ \hline \hline
    \multirow{4}{*}{$f_r$} & \multirow{3}{*}{LEO} & Lab1 $\to$ Lab1 & $1-3.22\times 10^{-5}$ \\ \cline{3-4}
 & & Lab2 $\to$ Lab2 & $1-3.50\times 10^{-5}$ \\ \cline{3-4}
 & & Lab1 $\rightleftarrows$ Lab2 & $1-3.36\times 10^{-5}$ \\ \cline{2-4}
 & VLEO & Lab1 $\to$ Lab1 & $1-4.71\times 10^{-5}$ \\ \hline
  \multirow{3}{*}{$f_l$} & \multirow{2}{*}{LEO} & Sat $\rightleftarrows$ Lab1 & $1 \pm 1.61\times 10^{-5}$ \\ \cline{3-4}
 & & Sat $\rightleftarrows$ Lab2 & $1 \pm 1.75\times 10^{-5}$ \\ \cline{2-4}
 & VLEO & Sat $\rightleftarrows$ Lab1 & $1 \pm 2.35\times 10^{-5}$ \\ \hline
\end{tabular}
\caption{Dimensionless physical parameters present in the expression of the energy shifts, with the corresponding shifts. Different signs correspond to different configurations: upper signs refer to downlinks, while lower signs denote uplinks.}
\label{paramtable}
\end{table}
The numerical values for the rescaled angular constants $l_{\phi}$ and $\kappa$ are obtained through the approximate analytical expression \eqref{azimuthal_cst_pert} and by numerical integration of the constraint equation \eqref{Carter_eq_gen_constraint_pert}, respectively. The following numerical values were used to obtain these parameters: we considered two stations on Earth with angular coordinates $(\theta_1, \phi_1)=(37.48^{\circ},13.40^{\circ})$ and $(\theta_2,\phi_2)=(51.88^{\circ},13.36^{\circ})$, namely in Berlin (Germany) and Palermo (Italy) respectively. We denote these stations by Lab1 and Lab2, respectively. The reflection or the measurement aboard the satellite occur at $(\theta_s,\phi_s)=(15^{\circ},13.38^{\circ})$ for LEO orbits or $(\theta_s,\phi_s)=(30^{\circ},13.38^{\circ})$ for VLEO orbits, depending on the orbit considered. Such coordinates guarantee the validity of the approximate expression \eqref{azimuthal_cst_pert} for the azimuthal angular momentum constant $l_{\phi}$. Notice that in the case of VLEO orbits, there is no null geodesic that connects the satellite to Lab2. For a scheme using a VLEO orbit, we will therefore only consider light rays exchanged with Lab1. 

We then proceed to compute the angular parameter $\delta_{\text{ang}}$ using the previously obtained parameters together with the choice $\epsilon_{\omega}=\epsilon_{\zeta}=1$ in Eq. \eqref{angular_function}. Finally, we are able to obtain numerical values for the frequency shifts \eqref{energy_reflect_exact} and \eqref{freq_shift_exact}, which are displayed in the second part of Table \ref{paramtable}.

\subsubsection{Precision bounds results}\label{bounds_values}
We are finally in the position of employing \eqref{rs_bound} and \eqref{QFI0_expl} to give numerical values for the highest sensitivity attainable in the measurements discussed above, using (V)LEO satellites. We use the numerical values given in Sec. \ref{param_values} together with the number of probes of $N=10^{10}$, the bandwidth in frequencies of $\sigma=10^6$ Hz and the peak frequency of $\Omega_0=7\times10^{14}$ Hz. Table \ref{precisiontable} contains our final estimates.

\begin{table}[!htbp]
\begin{tabular}{| c | c | c | c | c |}
\hline
\multirow{2}{*}{Orbit} & Light's & \multirow{2}{*}{$\Delta r_S/r_S$} & \multirow{2}{*}{$\Delta R_E/R_E$} & \multirow{2}{*}{$\Delta h/h$} \\
 & trajectory & & & \\ \hline
\multirow{5}{*}{LEO} & Lab1 $\leftrightarrows$ Lab2 & 8.50 $\times10^{-10}$ & 1.12 $\times10^{-9}$ & 3.56 $\times10^{-9}$ \\ \cline{2-5}
 & Lab1 $\to$ Lab1 & 8.87 $\times10^{-10}$ & 1.17 $\times10^{-9}$ & 3.71 $\times10^{-9}$ \\ \cline{2-5}
 & Lab2 $\to$ Lab2 & 8.15 $\times10^{-10}$ & 1.07 $\times10^{-9}$ & 3.41 $\times10^{-9}$ \\ \cline{2-5}
 & Sat $\to$ Lab1 & 1.77 $\times10^{-9}$ & 2.33 $\times10^{-9}$ & 7.43 $\times10^{-9}$ \\ \cline{2-5}
 & Sat $\to$ Lab2 & 1.63 $\times10^{-9}$ & 2.14 $\times10^{-9}$ & 6.83 $\times10^{-9}$ \\ \hline
 \multirow{2}{*}{VLEO} & Lab1 $\to$ Lab1  & 6.06 $\times10^{-10}$ & 6.30 $\times10^{-10}$ & 1.57 $\times10^{-8}$ \\ \cline{2-5}
 & Sat $\to$ Lab 1 & 1.21 $\times10^{-9}$ & 1.26 $\times10^{-9}$ & 3.15 $\times10^{-8}$ \\ \hline
\end{tabular}
\caption{Precision bounds obtained through the quantum metrology scheme described above, for the different possible configurations of the reflecting and downlink schemes. Results for uplinks are very similar to those in downlinks.}
\label{precisiontable}
\end{table}

In the case of reflecting schemes in Sec. \ref{reflecting_scheme}, irrespectively of the orbit and configuration considered, we obtain a bound for the relative error on the measurement of the Earth's Schwarzschild radius of the order $\Delta r_S/r_S\sim 10^{-10}$. The state-of-the-art in classical experiments gives a value of $2\times10^{-9}$ \cite{IAG1999}. Without taking noise into account, our result would therefore be one order of magnitude better than the current state of the art using classical measurements. The results obtained for the link scheme in Sec. \ref{link_scheme} are a great improvement from previous works considering squeezed states of photons sent radially from Earth and measured by a satellite \cite{David2,Kerr_sat}. The precision obtained is now of the order of magnitude of the state-of-the-art measurement, namely $\sim10^{-9}$. Note that we could have used squeezed states in order to improve the result further. We leave further analysis in this direction to future work.

We also find bounds on relative errors for the Earth's average local radius (over the area on Earth spanned by the light ray's propagation above it), which are of the order $\Delta R_E/R_E\sim10^{-9}-10^{-10}$ for different configurations involving the LEO polar satellite and VLEO respectively. For a measurement of the altitude of the satellite, the result is better for the LEO set ups with a precision bound of $\Delta h/h\sim10^{-9}$, while we have $\Delta h/h\sim10^{-8}$ for the VLEO schemes.

Note that in the reflecting scheme, for all the configurations we considered we had $\epsilon_{\theta}'=-\epsilon_{\theta}$. It is particularly advantageous to work in such configurations since the leading terms in the overlap \eqref{momentum_overlap_reflect} add up instead of competing. The effect is therefore significantly larger for such configurations than for set ups with $\epsilon_{\theta}'=\epsilon_{\theta}$. Also note that despite the better results of the reflecting scheme, the link schemes provide a channel that crosses the atmosphere only once, while in the reflecting scheme the light needs to cross it twice. Therefore, one should expect much less atmospheric noise in the link schemes in Sec. \ref{link_scheme}.

Finally we notice that here, contrarily to the scheme studied in previous work \cite{Kerr_sat}, the result obtained is far better for LEO satellites than for GEO satellites. There are two reasons for this. The first one is that we considered schemes where most of the angular momentum of the photon is polar and not azimuthal, while a GEO satellite only has an azimuthal velocity. This means that the reflection by a GEO satellite cannot transfer much energy to the photons in the reflecting scheme. In the link scheme, the change in energy due to the angular nature of the beam appears through the Doppler shift occurring because of the satellite's motion. Because the light's angular momentum is mostly polar, and because the GEO satellite only has an azimuthal velocity which is also quite slow compared to satellites in LEO, the resulting Doppler shift is very small when using GEO. Instead, the polar satellites in LEO and almost polar VLEO that we considered see much larger Doppler shifts, also because of their greater velocities.  The second reason is due to the very nature of the effect: in previous work \cite{Kerr_sat}, both gravitational shift and Doppler shift of frequencies are measured, and the latter effect dominates for a LEO orbit, but it is reduced in part by the competing gravitational redshift. Furthermore, for photons sent to a GEO satellite, the gravitational redshift largely dominates, which explains the stronger result than for a LEO orbit in the radial link scenarios \cite{Kerr_sat}. However, in the schemes we study here, the effect comes either from the satellite directly communicating its kinetic energy to the reflected photons in the scheme in Sec. \ref{reflecting_scheme}, or from the angularity of the beam in the link schemes in Sec. \ref{link_scheme}. These are significantly stronger effects, especially for lower orbiting satellites which have larger velocities. 

\section{Quantum bit error rate}\label{QBER_section}

In this section we briefly compute the QBER for a simple QKD protocol implemented in either the reflecting or the link scheme which we have previously introduced for metrology purposes. We can evaluate how the QKD protocol described in \cite{David1,Kerr_sat} is impaired by relativistic effects through the computation of the QBER, namely the rate of failure in the communication. We here find the following QBER:
\begin{align}\label{QBER_formula}
\text{QBER}=&\, 1-\mathcal{F}(\ket{\boldsymbol{\gamma_{0}}},\ket{\boldsymbol{\gamma_{\epsilon}}})  \\
\approx&\, 1-e^{-\left[\delta_{\text{ang}}\left(\epsilon_{\theta},\kappa,l_{\phi}\right)-\delta_{\text{ang}}\left(\epsilon_{\theta}',\kappa ',l_{\phi}' \right) \right]^2 \frac{\Omega_0^2}{4 \sigma^2}\epsilon^2},
\end{align}
for the reflecting schemes in Sec. \ref{reflecting_scheme}. For the link schemes in Sec. \ref{link_scheme}, Eq. \eqref{QBER_formula} holds but with $\delta_{\text{ang}}\left(\epsilon_{\theta}',\kappa ',l_{\phi}' \right)=0$. Using the parameters given in Sec. \ref{param_values}, we can compute the QBER for different configurations of the link and reflecting schemes. For all the configurations studied for metrology purposes, for which we gave precision results in Table \ref{precisiontable}, we find a QBER approaching 100\%. Namely, the relativistic effects are so important that without properly correcting for them one wouldn't be able to use such configurations for establishing a QKD protocol. These effects can be canceled by shifting the frequency distribution of the fiducial state at measurement by the relevant amount given in Sec. \ref{frequency_shifts}. Note that polarization-based communications are completely resilient to the effect described here, since they affect only the frequency modes.

\section{Conclusion}\label{conclusion}
In this work we provided the theory for realistic Earth-to-satellite photon exchange schemes that can potentially lead to improved measurements of spacetime parameters of the Earth, such as the Schwarzschild radius. We employed quantum metrology techniques to estimate physical parameters of the spacetime of the Earth that are encoded into quantum superpositions of momentum-helicity states of photons traveling between Earth and a satellite. In the regimes considered here, we found that the effect from the Wigner phases acquired by the helicity states \cite{helicity} are negligible for quantum metrology purposes compared to the changes experienced by the momentum states.

Our results depend on the type of orbit of the satellite. For a satellite in geostationary (GEO) orbit, the photons are little affected by the relativistic effects. On the contrary, when a satellite is located in low Earth orbit (LEO), its velocity is much larger and therefore becomes important. The quantum states of the photons are affected by the change in energy imparted by the satellite in the reflecting scheme, which results in significant changes of the state. In the link scheme, the high velocity of the emitting (respectively receiving) satellite leads to large Doppler shifts in the photons' states. Comparing the received state with a fiducial state, we can obtain a bound on the precision for the quantum measurement of the Schwarzschild radius of the Earth. Given realistic parameters, our work shows that the schemes considered have the potential of outperforming measurements obtained with the state-of-the-art through classical means. We also find good precisions for the bound on the measurement of the Earth's radius and of the satellite's altitude. We believe that our results show a potential for improved parameter estimation. However, noise was not considered in this study, and a more realistic study could be made following the derivation of \cite{losses}. We leave this to future work.

The increased precision obtained from the measurements proposed here would have an immediate implication on the development of space-based quantum technologies as measuring instruments. A better accuracy for the values of the Earth's physical parameters and of other parameters, such as the altitude of the satellite, would have a direct effect on global navigation satellite systems (GNSS) and on other planned satellite-based quantum networks. Our study shows that the use of quantum-metrology targeting inherent quantum features of relativistic and quantum systems can open the door to conceptually novel designs of future technologies. 

\section*{Acknowledgments}
We thank J. Louko, L. Mazzarella, and D. Oi for useful comments and discussions. This work was in part supported by the Anglo-Austrian Society. D.E.B. acknowledges the COST Action CA15117 and CA15220 for partial support. J.K. thanks the University of Vienna for hospitality.

\appendix

\bibliographystyle{apsrev4-1}
\bibliography{reflect}

\end{document}